\documentclass[11pt,a4paper]{article}
\usepackage{sapem2005_old}
\usepackage[dvips]{graphicx}
\usepackage{times}
\usepackage[]{fontenc}
\usepackage[latin1]{inputenc}
\usepackage{latexsym,amssymb,amsmath}
\input epsf
 \usepackage[francais]{babel}
 \catcode`\é=13
 \defé{\'e}
 \catcode`\è=13
 \defè{\`e}
 \catcode`\à=13
 \defà{\`a}
 \catcode`\ê=13
 \defê{\^e}
 \catcode`\ù=13
 \defù{\`u}
 \catcode`\û=13
 \defû{\^u}
 \catcode`\ç=13
 \defç{\c c}
 \catcode`\â=13
 \defâ{\^a}
 \catcode`\î=13
 \defî{\^i}
 \catcode`\ô=13
 \defô{\^o}
\begin{document}
\pagestyle{empty}
%

%
\papertitle{Acoustic identification of a poroelastic cylinder}

\listauthors{Zine Fellah$^{1}$, Jean-Philippe Groby$^{1,2}$, Erick
Ogam$^{1}$, Thierry Scotti$^{1}$ et Armand Wirgin$^{1}$}

\listaddress{$^{1}$ Laboratoire de Mécanique et d'Acoustique, (UPR
7051 du CNRS), 13402 Marseille cedex 20, France\\
$^{2}$Laboratorium voor Akoestiek en Thermische Fysica,
 Katholieke Universiteit Leuven, 3001 Heverlee, Belgium}

\abstract{We show how to cope with the acoustic identification of
poroelastic materials when the specimen is in the form of a
cylinder.  We apply our formulation, based on the Biot model,
approximated by the equivalent elastic solid model, to a long
bone-like or borehole sample specimen probed by low frequency
sound.}
\section{Introduction}\label{intro}
It has become fairly common \cite{lapa84}, \cite{layo86},
\cite{asco87}, \cite{wesa89}, \cite{mcpa91}, \cite{wi92},
 \cite{hoot97}, \cite{kiiy95b}, \cite{drbe98},
\cite{we00}, \cite{fefe03a}, \cite{bugi03}, \cite{femi03},
\cite{febe04d}, \cite{bugi04},  to acoustically identify the
structural/material properties, and/or the phase
velocity/attenuation in poroelastic (e.g., biological
 or geophysical) materials by
processing data relative to the reflected and/or transmitted
pulses of a slab-like specimen of the material in response to an
incident plane-wave pulse. In most of these studies, the model of
the medium is that of a fluid or (less often)  viscoelastic solid.
It is not often that materials, especially those of biological
\cite{khvi81} or geophysical \cite{zale04} nature, present
themselves in the form of slabs, plates or layers, nor is it
judicious to cut them (an operation that is rarely accurate and
which often modifies the physical properties) to fit this shape.
We show herein how to cope with the acoustic identification of
these materials, in the form of cylindrical specimens
\cite{khvi81}, \cite{zale04}, whether they are considered to be
fluid-like or elastic solid-like. Work on the fully-poroelastic
(in the sense of Biot) case is in progress.
\newline
\newline
More specifically, this investigation is concerned with the
reconstruction of the material constants
$\lambda^{1}=1/\kappa^{1}$ ($\kappa^{1}$ the complex
compressibility), $\mu^{1}$ (shear modulus), $\rho^{1}$ (density)
of an elastic solid-like (assumed to be of this nature even if the
target is poroelastic in the sense of Biot), almost-circular
cylinder, modeled as a circular cylinder of radius $a$ close to
the average radius of the almost-circular cylindrical target. The
latter is immersed in a fluid-like host wherein propagates
plane-wave like acoustic probe radiation. The action of this wave
on the target results in a scattered acoustic wavefield which
serves to reveal the material properties of the target.
\newline
\newline
The wavelength ($\Lambda_{L}^{0}=2\pi/k_{L}^{0}$) of the probe
radiation is assumed to be much larger
($\chi_{L}^{0}:=k_{L}^{0}a<<1$)  than $a$. A perturbation analysis
is shown to enable an explicit reconstruction of $\rho^{1}$ and of
a linear combination of $\lambda^{1}$, $\mu^{1}$.
\newline
\newline
This reconstruction technique relies on the a priori knowledge of
$a$; it is assumed herein that this parameter is known.
\section{Physical configuration}\label{s2}
The scattering body is an infinite cylinder whose generators are
parallel to the $z$ axis in the cylindrical coordinate system
$(r,\theta,z)$. The intersection of the cylinder, within which is
located the origin $O$, with the $xOy$ plane defines:
\newline
\newline
i) the boundary curve $\Gamma=\{r=f(\theta); 0\le \theta < 2\pi
\}$, with $f$ a continuous, single-valued function of $\theta$;
further on, we shall take $\Gamma$ to be a circle, i.e.,
$f(\theta)=a$, with $a$ its radius, close to the average value of
$\eta(\theta)$,
\newline
\newline
ii) the bounded (inner) region (i.e., the one occupied by the body
in its cross-section plane) $\Omega_{1}=\{r<\eta(\theta);~0\le
\theta < 2\pi \}$,
\newline
\newline
iii) the unbounded (outer) region $\Omega_{0}=\{r>\eta(\theta);~
0\le \theta < 2\pi \}$.
\newline
\newline
It is assumed that $\Omega_{0}$ is filled with a linear,
homogeneous, inviscid fluid  $M^{0}$ and $\Omega_{1}$ by a linear,
macroscopically-homogeneous, isotropic,
 porous medium $M^{1}$ which will subsequently be
associated with a linear, homogeneous, isotropic, time-invariant
elastic solid medium.
\newline
\newline
The material constants of $M^{0}$ are assumed to be known. Those
of $M^{1}$ are unknown and are to be recovered by the technique
described hereafter. The latter relies on probing the cylinder
(from the outside) by a plane acoustic wave whose wavevector lies
in the $xOy$ plane.
\section{Ingredients of the Biot theory}
We give the ingredients of the  basic Biot theory of biphasic
(solid/fluid) porous media \cite{gusc99}, \cite{ya83}.
\subsection{Conservation of momentum relations}
In the absence of applied body forces, the conservation of
momentum relations  take the form
\newline
\begin{equation}\label{gusc1}
\nabla\cdot\boldsymbol{\sigma}-\rho\mathbf{u}-\rho_{f}\mathbf{w}_{,tt}=\mathbf{0}~~~,~~~
\nabla p+\rho_{f}\mathbf{u}_{,tt}+m\mathbf{w}_{,tt}+\frac{\eta F
}{\kappa}\mathbf{w}_{,t}=0~,
\end{equation}
\newline
wherein:
\newline
\newline
- $\mathbf{f}_{,t}$ designates a first-order partial derivative
with respect to time $t$, and $\mathbf{f}_{,tt}:=(f_{,t})_{,t}$,
\\
- $\boldsymbol\sigma$ is the total stress tensor in the porous
fluid-saturated medium,
\\
- $\rho_{s}$ the density of the solid component
\\
- $\rho_{f}$  the density of the viscous fluid filling the
(interconnected) pores,
\\
- $\rho$  the bulk density of the porous medium, such that
$\rho=(1-\phi)\rho_{s}+\phi\rho_{f}$, with $\phi$ the porosity
(volume fraction of fluid relative to total volume in a
representative volume element),
\\
- $\boldsymbol{\sigma}$ the stress tensor,
\\
- $\mathbf{u}$ the displacement vector of the  solid particle
component,
\\
- $\mathbf{U}$  the fluid particle displacement vector,
\\
- $\mathbf{w}$ the relative displacement  vector of the fluid
particle relative to the solid particle (both particles assumed to
occupy the same point) defined as
$\mathbf{w}:=\phi(\mathbf{U}-\mathbf{u})$,
\\
- $p$  the pressure in the fluid component of the porous medium,
\\
- $\eta$  the viscosity of this fluid,
\\
- $\kappa$  the (low-frequency) permeability,
\\
- $m=\frac{\rho_{f}\alpha}{\phi}$ the virtual mass,
\\
- $\alpha$ the tortuosity  (in \cite{ch95} $\alpha$ is termed the
virtual mass coefficient or structure factor), which, in
\cite{be81}, is related to the porosity by $\alpha=1+r\left(
\frac{1-\phi}{\phi}\right) $ (wherein $r$ is a constant with a
value between 0 and 1),
\\
- $F(t)$ is a linear integral convolution operator with respect to
time  which, in the frequency domain, becomes a
frequency-dependent multiplier $F(\omega)$, implying
frequency-dependent permeability, i.e.,
$\tilde{\kappa}(\omega)=\frac{\kappa}{F(\omega)}$, wherein
$\tilde{\kappa}(\omega)$ is the so-called dynamic permeability,
and $F$ is designed so that $\lim_{\omega\rightarrow
0}F(\omega)=1$.
\subsection{Constitutive relations}
Biot's constitutive relations \cite{gusc99},\cite{stka81}
 linearly relate the total stress and fluid pressure to
the (isotropic) solid and fluid displacement spatial derivatives
via
\newline
\begin{equation}\label{gusc23}
  \boldsymbol{\sigma}=2\mu\boldsymbol{\varepsilon}+\left[ (H-2\mu)e-C\zeta\right]
  \mathbf{I}~~~,~~~
  p= M\zeta-Ce~
\end{equation}
\newline
with:
\newline
\begin{equation}\label{gusc25}
D=K_{s}\left[1 + \phi\left( \frac{K_{s}}{K_{f}}-1\right)
 \right] ~~~,~~~H=\frac{(K_{s}-K)^{2}}{D-K}+K+\frac{4\mu}{3}~,
\end{equation}
\newline
\begin{equation}\label{y8}
C=\frac{K_{s}(K_{s}-K)}{D-K}~~~,~~~M=\frac{K_{s}^{2}}{D-K}~~~,~~~
e=\nabla\cdot\mathbf{u}~~,~~\zeta=-\nabla\cdot\mathbf{w}~,
\end{equation}
\newline
wherein:
\newline
\newline
- $\boldsymbol{\varepsilon}$ is the strain tensor
$\boldsymbol{\varepsilon}=\frac{1}{2}\left(
\nabla\mathbf{u}+\nabla\mathbf{u}^{T}\right) $,
\\
- $\mathbf{I}$  the unit tensor
\\
- $\mu$ the shear modulus (rigidity) of the saturated solid,
\\
- $K=(\lambda+\frac{2}{3}\mu)$, $K_{s}$  the bulk moduli of the
dry (i.e., drained) solid matrix and solid grain material
respectively (note that  $\lambda_{c}=\left(
1-\frac{K}{K_{s}}\right) C+\lambda$ is the Lam\'e constant of the
saturated solid),
\\
- $K_{f}$ the fluid bulk modulus.
\subsection{Equations of motion in terms of $\mathbf{u}$ and
$\mathbf{w}$}\label{gusceqmot}
We assume henceforth that all the material parameters are
constants with respect to position, i.e., {\it the medium is
macroscopically homogeneous}.
The conservation of momentum equations and constitutive relations
are employed in such a way as to eliminate the pressure and stress
tensor so as to obtain:
\newline
\begin{equation}\label{gusc52y}
 \mu\nabla^{2}\mathbf{u} +(H-\mu)\nabla e-
  C\nabla
 \zeta=\rho \mathbf{u}_{tt}+\rho_{f}\mathbf{w}_{,tt}~~~,~~~
  C\nabla e-M \nabla\zeta=\rho_{f}\mathbf{u}_{,tt}+m \mathbf{w}_{,tt}+
  \frac{\eta}{\kappa}F\mathbf{w}_{,t}~,
\end{equation}
\newline
which is the vectorial form of the Biot wave equations as given by
Yamamoto \cite{ya83} for a macroscopically-homogeneous porous
medium.
\section{Choice of an approximate model to describe wave
propagation in a porous medium}
The fundamental difficulty with the Biot theory is twofold: i)
{\it two} coupled (vectorial) wave equations have to be solved
simultaneously, and ii) many material parameters have to be
recovered in the inverse problem context. This is why the
traditional approach (notably in the underwater acoustics
community) has been to reduce this model to a simpler one (with
fewer material parameters) in which only one (vectorial or scalar)
wave equation has to be solved.
\subsection{Equivalent elastic solid model (EESM) obtained from the
limit $\phi\rightarrow 0$}
The introduction of the second equation of (\ref{gusc52y}) into
the first gives
\newline
\begin{equation}\label{gusc54y}
 \mu\nabla^{2}\mathbf{u} +\left( H-\mu-\frac{C}{m}\right) \nabla e+\left( -
  C+\frac{M}{m}\right) \nabla
 \zeta+\left(-\rho+\frac{\rho_{f}}{m}\right) \mathbf{u}_{tt}+
 \frac{\eta}{m\kappa}F\mathbf{w}_{,t}=\mathbf{0}~,
\end{equation}
\newline
Recalling the definitions of $m$ and $\mathbf{w}$, we conclude
that $\lim_{\phi\rightarrow 0}~m^{-1}=0$ and
$\lim_{\phi\rightarrow 0}~\mathbf{w}=\mathbf{0}$, so that in the
limit $\phi\rightarrow 0$, (\ref{gusc54y}) becomes
\newline
\begin{equation}\label{gusc55y}
 \mu\nabla^{2}\mathbf{u} +\left( \lambda_{c}+\mu\right) \nabla \nabla\cdot \mathbf{u}-
  \rho\mathbf{u}_{tt}=\mathbf{0}~,
\end{equation}
\newline
which is simply the Navier wave equation in a non-dissipative,
linear, homogeneous, isotropic solid in which the material
parameters are $\lambda_{c},~\mu, ~\rho=\rho_{s}$.
\newline
\newline
This equation forms the basis of the so-called equivalent elastic
solid model for wave propagation in poroelastic media, often
employed for the evaluation of transmission loss of sound over
sediment layers on sea bottoms \cite{bugi98}, \cite{huel90}.
\subsection{Equivalent viscoelastic solid model (EVSM) for wave
propagation in poroelastic media}
We now consider another approximation of the Biot wave equations
which leads to  what has been termed the equivalent viscoelastic
solid model of propagation in poroelastic media
\cite{moba96}\cite{stka81}.
\newline
\newline
The basic idea is to reduce the Biot model to only one wave
equation, while retaining the loss mechanism inherent in this
model (contrary to what is done in the equivalent elastic solid
model in which the loss mechanism is abolished in the limit
$\phi\rightarrow 0$).
\newline
\newline
Let us return to the definition of $\mathbf{w}$, i.e.,
$\mathbf{w}=\phi(\mathbf{U}-\mathbf{u})$ in which we assume
$\phi\mathbf{U}<<\phi\mathbf{u}$. Thus, neglecting the terms in
$\phi\mathbf{U}$ in (\ref{gusc54y}), we get
\newline
\begin{equation}\label{gusc56y}
 \mu\nabla^{2}\mathbf{u} +\left[ H-\mu-\frac{C}{m}-C\phi+\frac{M\phi}{m}\right]
 \nabla\nabla\cdot \mathbf{u}+\left( -\rho+\frac{\rho_{f}}{m}\right)
 \mathbf{u}_{tt}-
 \frac{\eta\phi}{m\kappa}F\mathbf{u}_{,t}=\mathbf{0}~.
\end{equation}
\newline
This wave equation for the displacement in the solid component is
similar to the Navier equation, with the exception that the
wavenumber is now complex, its real part not being equal to that
of the Navier equation wavenumber, and its imaginary part being
conditioned by $\frac{\eta\phi}{m\kappa}F$.
\subsection{Equivalent fluid model (EFM)}\label{efm}
The equivalent fluid model (EFM) is  appropriate when the fluid is
light (e.g., a gas such as air) and the solid skeleton is
therefore relatively immobile (i.e., rigid). This model has been
employed, even when the fluid is not light, notably in the
underwater acoustics community \cite{vi80}, \cite{bugi98},
\cite{ya83}.
\newline
\newline
There exist various versions of the EFM, several of which are
described and compared in Depollier {\it et al.} \cite{deal88}.
The one we shall consider herein is a simplified version of the
model offered in  \cite{fefe03a}.
\newline
\newline
No restrictions are introduced concerning the porosity, but, for
the sake of simplicity, the fluid viscosity is assumed to be nil.
The fundamental assumption is that the solid component is rigid,
i.e.
\begin{equation}\label{efm2}
K_{s}=\infty,~~~,~~~\mathbf{u}=0~~\rightarrow \mathbf{u}_{,tt}=0~.
\end{equation}
\newline
If, in addition, the medium is macroscopically-homogeneous and
time-invariant (i.e., $\rho_{f}$, $\alpha$ and $K_{f}$ are
constants with respect to position and $t$) then the Biot system
of equations reduce to
\newline
\begin{equation}\label{efm12}
-\nabla^{2} p- \rho_{f}\alpha \nabla\cdot\mathbf{U}_{,tt}=0
 ~~~,~~~
  p_{,tt}+ K_{f}\mathbf{U}_{,tt}=0~,
\end{equation}
\newline
a linear combination of which yields
\newline
\begin{equation}\label{efm14}
 \nabla^{2} p- \frac{\alpha\rho_{f}}{K_{f}}p_{,tt}=0~,
\end{equation}
\newline
which is the wave equation in the equivalent fluid.
\newline
\newline
Note that $\alpha=1$ for a homogeneous fluid.
\subsection{Our choice of approximate model}
Since all three approximate models reduce to the same type of wave
equation (either vectorial or scalar), a generic
choice--(vectorial in nature) adopted herein--is the EESM.
\newline
\newline
This means that we replace, by thought, the porous medium cylinder
immersed in an inviscid fluid by an elastic solid cylinder
immersed in the same fluid.
\section{Mathematical description of the problem}
%
\subsection{Preliminaries}
 Due to the invariance of the cylinder and incident
wavefield $p^{i}$ with respect to $z$, the incident and scattered
fields are also invariant with respect to $z$.
\newline
\newline
Let $p^{0}$ designate pressure in $\Omega_{0}$; due to this
invariance, $p^{0}=p(x,y,t)=p^{i}(x,y,t)+p^{d}(x,y,t)$, wherein
$p^{d}$ is the diffracted pressure in $\Omega_{0}$.
\newline
\newline
For the same reason, the total displacement wavefield $\mathbf{u}$
in $\Omega_{1}$ is of the form $\mathbf{u}=\mathbf{u}(x,y,t)$.
\newline
\newline
The analysis is carried out in the space-frequency domain via
$p^{0,i,d}(\mathbf{x},t)=\int_{-\infty}^{\infty}p^{0,i,d}(\mathbf{x},\omega)\exp(-i\omega
t)d\omega$ and
$\mathbf{u}^{1}(\mathbf{x},t)=\int_{-\infty}^{\infty}\mathbf{u}^{1}(\mathbf{x},\omega)\exp(-i\omega
t)d\omega$ wherein $\mathbf{x}=(x,y)$.
\newline
\newline
Henceforth, it is implicit that $p^{0,i,d}$ (meaning $p^{0}$,
$p^{i}$ or $p^{d}$) and $\mathbf{u}^{1}$ are functions of
$(\mathbf{x},\omega)$.
\subsection{Governing equations}
The incident plane wave is
\newline
\begin{equation}\label{2.6}
p^{i}=S(\omega)\exp\left( -ik_{0}r\cos\left(
\theta-\theta^{i}\right) \right)
\end{equation}
\newline
(wherein $S(\omega)$ is the amplitude spectrum, $\theta^{i}$ the
incident angle in the $xOy$ plane, and $k^{0}=\omega/c^{0}$, with
$c^{0}=(\rho^{0}\kappa^{0})^{-1/2}$).
\newline
\newline
$p^{0,i,d}$ satisfy the frequency-domain pressure wave (Helmholtz)
equation
\newline
\begin{equation}\label{2.7}
\left( \nabla^{2} +(k_{L}^{0})^{2}\right)
p^{0,i,d}=0\text{~~~~~in~ }\Omega_{0}~,
\end{equation}
\newline
and the radiation condition
\newline
\begin{equation}\label{2.8}
p^{d}_{,r}-ik^{0}p^{d}=o(r^{-1/2})~;~r\rightarrow\infty~,~
\forall\theta\in[0,2\pi[ ~.
\end{equation}
\newline
$\mathbf{u}^{1}$ satisfies the frequency-domain elastic wave
equation
\newline
\begin{equation}\label{2.9}
\mu\nabla^{2}\mathbf{u}^{1}+(\lambda^{1}+\mu^{1})\nabla\cdot\nabla\mathbf{u}^{1}+\omega^{2}\mathbf{u}^{1}=0
\text{~~~~~in~}\Omega_{1}~,
\end{equation}
\newline
and the boundedness condition
\newline
\begin{equation}\label{2.10}
\|u^{1}\|<\infty~  \text{~~~~~in~}\Omega_{1}~.
\end{equation}
\newline
Let $\boldsymbol{\nu}$ designate the unit outward-pointing (from
$\Omega^{1}$) unit normal vector,
$\mathbf{T}^{j}=\boldsymbol{\sigma}^{j}\cdot\boldsymbol{\nu}$ the
traction. Then the transmission boundary conditions are:
\newline
\begin{equation}\label{2.11}
\mathbf{T}^{0}-\mathbf{T}^{1}=\mathbf{0}~~~,~~~
\mathbf{u}^{0}\cdot\boldsymbol{\nu}-\mathbf{u}^{1}\cdot\boldsymbol{\nu}=0~\text{~~~~~on~
} \Gamma~.
\end{equation}
\subsection{Reduction of the elastic solid wave equation to two
Helmholtz equations}
The use of the Helmholtz decomposition
\newline
\begin{equation}\label{h1}
\mathbf{u}^{1}=\nabla\varphi^{1}+\nabla\times\boldsymbol{\psi}^{1}~,
\end{equation}
enables (\ref{2.9}) to be reduced to the two (one scalar, the
other vectorial) Helmholtz equations
\begin{equation}\label{h2}
\left( \nabla^{2}+(k_{L}^{1})^{2}\right)\varphi^{1}=0~~~,~~~\left(
\nabla^{2}+(k_{T}^{1})^{2}\right)\boldsymbol{\psi}^{1}=\mathbf{0}~,
\end{equation}
\newline
wherein
\newline
\begin{equation}\label{h3}
k_{L}^{1}=\frac{\omega}{c_{L}^{1}}=\omega\left(\frac{\lambda^{1}+2\mu^{1}}{\rho^{1}}\right)
^{-1/2}~~~,~~~k_{T}^{1}=\frac{\omega}{c_{T}^{1}}=\omega\left(\frac{\mu^{1}}{\rho^{1}}\right)
^{-1/2}~.
\end{equation}
\newline
Recalling that the fields $p$ and $\mathbf{u}$ do not depend on
$z$ enables (\ref{2.7}) and (\ref{h2}) to be cast into the
cylindrical coordinate forms:
\newline
\begin{multline}\label{h4}
~~~~~~~~~~~~~~p^{0}_{,rr}+r^{-1}p^{0}_{,r}+r^{-2}p^{0}
_{,\theta\theta}+(k_{L}^{0})^{2}p^{0}=0~~~,~~~~
\varphi^{1}_{,rr}+r^{-1}\varphi^{1}_{,r}+r^{-2}\varphi^{1}_{,\theta\theta}+(k_{L}^{1})^{2}\varphi^{1}=0
\\
\boldsymbol{\psi}^{1}_{,rr}+r^{-1}\boldsymbol{\psi}^{1}_{,r}+r^{-2}\boldsymbol{\psi}^{1}_{,\theta\theta}+
(k_{T}^{1})^{2}\boldsymbol{\psi}^{1}=0~.
~~~~~~~~~~~~~~~~~~~~~~~~~~~~~~~~~~~~~~~~~~
\end{multline}
\newline
The gauge condition $\nabla\cdot\boldsymbol{\psi}^{1}=0$ and the
absence of shear stress in the fluid imply that $\psi^{1}_{r}=
\psi^{1}_{\theta}=0$. In the cylindrical coordinate system, the
traction and normal component of displacement continuity
conditions reduce to:
\newline
\begin{multline}\label{h5}
~~~~~~~~~~~~~~~~~~~~~~~~~~~~~~~~~~~~~~
-p^{0}+\lambda^{1}(k_{L}^{1})^{2}\varphi^{1}-2\mu^{1}\left[
\varphi^{1}_{,rr}-r^{-2}\psi^{1}_{z,\theta}+r^{-1}\psi^{1}_{z,r\theta}\right]
=0 ~,
\\
~~~~~2\left[
-r^{-2}\varphi^{1}_{,\theta}+r^{-1}\varphi^{1}_{,r\theta}\right]
+\left[
-\psi^{1}_{z,rr}+r^{-1}\psi^{1}_{z,r}+r^{-2}\psi^{1}_{z,\theta\theta}\right]
=0~,
\\
\frac{1}{\lambda^{0}(k_{L}^{0})^{2}}p^{0}_{,r}-\phi^{1}_{,r}-r^{-1}\psi^{1}_{z,\theta}=0~.
~~~~~~~~~~~~~~~~~~~~~~~~~~~~~~~~~~~~~~~~~~~~~~~~~~
\end{multline}
\subsection{Field representations}
The incident pressure field satisfies the periodicity condition
$p^{i}(r,-\theta+2\theta^{i},\omega)=p^{i}(r,\theta,\omega)$ and
the first of the Helmholtz equations in  (\ref{h4}) so that (also
on account of (\ref{2.6})
\newline
\begin{equation}\label{fr1}
p^{i}=\sum_{m=0}^{\infty}a_{m}\epsilon_{m}J_{m}(k_{L}^{0}r)\cos
m(\theta-\theta^{i})~~~,~~~\text{with}~~a_{m}=S(\omega)e^{-im\frac{\pi}{2}}~,
\end{equation}
\newline
wherein $J_{m}(~)$ is the $m$-th order Bessel function and
$\epsilon_{0}=1~,~\epsilon_{m>0}=2$.
\newline
\newline
The periodicity of $p^{i}$ entails
$p^{d}(r,-\theta+2\theta^{i},\omega)=p^{d}(r,\theta,\omega)$, so
that on account of the first of the Helmholtz equations in
(\ref{h4}) and the radiation condition
\newline
\begin{equation}\label{fr2}
p^{d}=\sum_{m=0}^{\infty}b_{m}\epsilon_{m}H_{m}(k_{L}^{0}r)\cos
m(\theta-\theta^{i})~,
\end{equation}
\newline
wherein $H_{m}(~)=H_{m}^{(1)}(~)$ is the $m$-th order Hankel
function of the first kind.
\newline
\newline
The periodicity of $p^{i}$ and $p^{d}$ also entails
$\phi^{1}(r,-\theta+2\theta^{i},\omega)=\phi^{1}(r,\theta,\omega)$,
so that on account of the second of the Helmholtz equations in
(\ref{h4}) and the boundedness condition
\newline
\begin{equation}\label{fr3}
\phi^{1}=\sum_{m=0}^{\infty}c_{m}\epsilon_{m}J_{m}(k_{L}^{1}r)\cos
m(\theta-\theta^{i})~.
\end{equation}
\newline
By means of any one of the transmission conditions (\ref{h5}), and
on account of the periodicity conditions satisfied by $p^{i}$,
$p^{d}$, and $\phi^{1}$, it is found that $\psi^{1}_{z}$ obeys the
relation
$\psi^{1}_{z}(r,-\theta+2\theta^{i},\omega)=-\psi^{1}_{z}(r,\theta,\omega)$
so that  on account of the third of the Helmholtz equations in
(\ref{h4}) and the boundedness condition
\newline
\begin{equation}\label{fr4}
\psi^{1}_{z}=\sum_{m=0}^{\infty}d_{m}\epsilon_{m}J_{m}(k_{T}^{1}r)\sin
m(\theta-\theta^{i})~.
\end{equation}
\section{Use of the transmission boundary conditions to obtain
$\{b_{m}\}$, $\{c_{m}\}$, $\{d_{m}\}$}
We employ the orthogonality relations
\newline
\begin{equation}\label{tbc1}
\int_{\theta^{i}}^{\theta^{i}+\pi}\cos m(\theta-\theta^{i})\cos
n(\theta-\theta^{i})\frac{d\theta}{\pi}=\frac{\delta_{mn}}{\epsilon_{m}}~~~,
~~~\int_{\theta^{i}}^{\theta^{i}+\pi}\sin m(\theta-\theta^{i})\sin
n(\theta-\theta^{i})\frac{d\theta}{\pi}=\delta_{mn}\frac{(1-\delta_{m0})}{2}
~,
\end{equation}
\newline
wherein $\delta_{mm}=1~,~\delta_{mn\neq m}=0$ to obtain:
\newline
\begin{equation}\label{tbc2}
\mathbf{P}_{0}\mathbf{q}_{0}=\mathbf{r}_{0} ~,
\end{equation}
\newline
in which
\newline
\begin{equation}\label{tbc3}
\mathbf{P}_{0}=
  \begin{pmatrix}
  \chi_{L}^{0}\dot{H}_{0}(\chi_{L}^{0}) & -\upsilon_{L}^{0}\chi_{L}^{1}\dot{J}_{0}(\chi_{L}^{1})
  \\\\
  -H_{0}(\chi_{L}^{0}) & \upsilon_{L}^{1}J_{0}(\chi_{L}^{1})-2\mu^{1}(k_{L}^{1})^{2}\ddot{J}_{0}(\chi_{L}^{1})
\end{pmatrix}
~~~,~~~ \mathbf{q}_{0}=
 \begin{pmatrix}
 b_{0}
 \\\\
c_{0}
\end{pmatrix}
~~~,~~~\mathbf{r}_{0}=
 \begin{pmatrix}
   -a_{0}\chi_{L}^{0}\dot{J}_{0}(\chi_{L}^{0})\\\\
  a_{0}J_{0}(\chi_{L}^{0})
\end{pmatrix}~,
\end{equation}
\newline
and
\begin{equation}\label{tbc5}
\mathbf{P}_{n}\mathbf{q}_{n}=\mathbf{r}_{n}~~;~~n=1,2,...~,
\end{equation}
\newline
in which
\newline
\begin{equation}\label{tbc6}
\mathbf{P}_{n}=
  \begin{pmatrix}
  \chi_{L}^{0}\dot{H}_{n}(\chi_{L}^{0}) & -\upsilon_{L}^{0}\chi_{L}^{1}\dot{J}_{n}(\chi_{L}^{1}) &
  -\upsilon_{L}^{0}nJ_{n}(\chi_{T}^{1}) \\\\
  -a^{2}H_{n}(\chi_{L}^{0}) & (\chi_{L}^{1})^{2}\left[ \lambda^{1}J_{n}(\chi_{L}^{1})-
  2\mu^{1}\ddot{J}_{n}(\chi_{L}^{1})\right] & 2\mu^{1}n\left[ J_{n}(\chi_{T}^{1})-
  \chi_{T}^{1}\dot{J}_{n}(\chi_{T}^{1})\right] \\\\
  0 & 2n\left[ J_{n}(\chi_{L}^{1})-\chi_{L}^{1}\dot{J}_{n}(\chi_{L}^{1})\right] &
  -(\chi_{T}^{1})^{2}\ddot{J}_{n}(\chi_{T}^{1})+\chi_{T}^{1}\dot{J}_{n}(\chi_{T}^{1})-
  n^{2}J_{n}(\chi_{T}^{1})\\
\end{pmatrix}
\end{equation}
\newline
\begin{equation}\label{tbc7}
\mathbf{q}_{n}=
 \begin{pmatrix}
 b_{n}
 \\\\
c_{n}
\\\\
d_{n}
\end{pmatrix}
~~~,~~~\mathbf{r}_{n}=
 \begin{pmatrix}
   -a_{n}\chi_{L}^{0}\dot{J}_{n}(\chi_{L}^{0})\\\\
  a_{n}a^{2}J_{n}(\chi_{L}^{0})\\\\
  0
\end{pmatrix}~,
\end{equation}
\newline
with $\chi_{L}^{j}=k_{L}^{j}a$, $\chi_{T}^{j}=k_{T}^{j}a$,~
$\upsilon_{L}^{j}=\lambda^{j}(k_{L}^{j})^{2}$,~
$\dot{g}(\varsigma)=dg/d\varsigma$,~
$\ddot{g}=d^{2}g/d\varsigma^{2}$, and for $n=1,2,....$.
\newline
\newline
In principle, the matrix equations (\ref{tbc2}) and (\ref{tbc5})
enable to determine the unknown coefficient vectors
$\mathbf{q}_{0}$ and $\mathbf{q}_{n}~;~n=1,2,...$, and thus to
solve the forward-scattering problem, notably for the prediction
of the scattered pressure field $p_{d}$ in the host fluid.
\newline
\newline
Rather than do this, and since we are more interested, in the
present context, in solving the {\it inverse-scattering problem}
of the reconstruction of $\lambda^{1}$, $\mu^{1}$, $\rho^{1}$, we
adopt a different strategy for determining $\mathbf{q}_{0}$ and
$\mathbf{q}_{n}~;~n=1,2,...$.
\section{Low-frequency approximation of the solution of the forward-scattering
problem in the region outside of the body}\label{s8}
We now define this new strategy.
\newline
\newline
We first note that:
\newline
\begin{equation}\label{8.1}
\chi_{L}^{1}=\frac{k_{L}^{1}}{k_{L}^{0}}\chi_{L}^{0}=\frac{c_{L}^{0}}{c_{L}^{1}}\chi~~~,
~~~\chi_{T}^{1}=\frac{k_{T}^{1}}{k_{L}^{0}}\chi_{L}^{0}=\frac{c_{L}^{0}}{c_{T}^{1}}\chi
~,
\end{equation}
\newline
wherein $\chi:=\chi_{L}^{0}$. We assume that $\chi$ is small
enough (i.e., $0<\chi<<1$) for it to be true that
$0<\chi_{L}^{1}<<1$ and $0<\chi_{T}^{1}<<1$, and employ a
perturbation scheme, based on the smallness of $\chi$ (which, to
the very least, implies very low frequencies $\omega$ and/or small
cylinder radius $a$), to solve the matrix equations.
\newline
\newline
Thus, the arguments of all the Bessel and Hankel functions
appearing in the expression of $\mathbf{P}_{n}$ are small, which
fact authorizes use to be made of the small-argument asymptotic
forms
\newline
\begin{equation}\label{8.2}
J_{m}(\xi)\sim \frac{1}{m!}\left( \frac{\xi}{2}\right)
^{m}~~;~~H_{0}^{(1)}(\xi)\sim
\frac{2i}{\pi}\ln\xi~~,~~,~~H_{m}^{(1)}(\xi)\sim
-\frac{i(m-1)!}{\pi}\left( \frac{\xi}{2}\right)
^{-m}~~;~~\xi\rightarrow 0~,~m=0,1,... ~,
\end{equation}
\newline
To do this in a systematic manner, we expand $\mathbf{P}_{m}$,
$\mathbf{q}_{m}$, and $\mathbf{r}_{m}$ in series of powers of
$\chi$:
\newline
\begin{equation}\label{8.3}
\mathbf{P}_{m}(\chi)=\sum_{j=0}^{\infty}P_{m}^{(j)}\chi^{j}~~;~~P_{m}^{(j)}:=\frac{1}{j!}\frac{\partial
^{j} }{\partial\chi^{j}}\mathbf{P}_{m}(\chi)\big| _{\chi=0}~~~,~~~
\mathbf{q}_{m}(\chi)=\sum_{n=0}^{\infty}\mathbf{q}_{m}^{(n)}\chi^{n}~,
\end{equation}
\newline
\begin{equation}\label{8.4}
\mathbf{r}_{m}(\chi)=\sum_{l=0}^{\infty}\mathbf{r}_{m}^{(l)}\chi^{l}~~;~~\mathbf{r}_{m}^{(l)}:=\frac{1}{l!}\frac{\partial
^{l} }{\partial\chi^{l}}\mathbf{r}_{m}(\chi)\big| _{\chi=0}~,
\end{equation}
\newline
which, ~after ~introduction ~into ~the matrix equation
$\mathbf{P}_{m}\mathbf{q}_{m}=\mathbf{r}_{m}$, yields (after
comparison of powers of $\chi$)~~
$\sum_{n=0}^{l}\mathbf{P}_{m}^{(l-n)}\mathbf{q}_{m}^{(n)}=\mathbf{r}_{m}^{(l)}~~;~~l=0,1,2,....$,
which defines the recursive scheme for the determination of
$\mathbf{q}_{m}^{l}$:
\newline
\begin{equation}\label{8.5}
\mathbf{q}_{m}^{(0)}=\left( \mathbf{P}_{m}^{(0)}\right)^{-1}
\mathbf{r}_{m}^{(0)}~~~,~~~ \mathbf{q}_{m}^{l}=\left(
\mathbf{P}_{m}^{(0)}\right)^{-1}\left[
\mathbf{r}_{m}^{(l)}-\sum_{n=0}^{l-1}\mathbf{P}_{m}^{(l-n)}\mathbf{q}_{m}^{(n)}\right]
~~;~~l =1,2,....~.
\end{equation}
\newline
After a series of algebraic manipulations, the following
asymptotic form of $b_{m}$ is found:
\newline
\begin{equation}\label{8.6}
b_{0}=b_{0}^{(2)}\chi^{2}+O(\chi^{4})~~~,~~~
b_{1}=b_{1}^{(2)}\chi^{2}+O(\chi^{4})~~~,~~~~b_{m>1}=O(\chi^{4})~~;~~\chi\rightarrow
0~.
\end{equation}
\newline
wherein
\begin{equation}\label{8.7}
b_{0}^{(2)}=a_{0}\left( \frac{-i\pi}{4}\right) \left[
\frac{\lambda^{1}+\mu^{1}-\lambda^{0}}{\lambda^{1}+\mu^{1}}\right]
~~~,~~~ b_{1}^{(2)}=a_{1}\left( \frac{-i\pi}{4}\right) \left[
\frac{\rho^{0}-\rho^{1}} {\rho^{0}+\rho^{1}}\right] ~,
\end{equation}
\newline
so that the diffracted pressure field in the host fluid becomes
(to second order in $\chi$)
\begin{equation}\label{8.8}
p^{d}(r,\theta,\omega)\approx
b_{0}^{(2)}\chi^{2}H_{0}^{(1)}(k_{L}^{0}r)+
2b_{1}^{(2)}\chi^{2}H_{1}^{(1)}(k_{L}^{0}r)\cos(\theta-\theta^{i})
~.
\end{equation}
\section{Explicit resolution of the inverse problem: recovery of
the material parameters of the specimen}
By making use of (\ref{8.8}) and the first of the orthogonality
relations (\ref{tbc1}) we find
\newline
\begin{equation}\label{8.9}
B_{m}:=\frac{4i}{\pi}\frac{b_{m}^{(2)}}{a_{m}}=\frac{4i}{\pi
a_{m}\chi^{2}}\frac{1}{H_{m}^{(1)}(k_{L}^{0}b)}
\int_{\theta^{i}}^{\theta^{i}+\pi}p^{d}(b,\theta,\omega)\cos m
(\theta-\theta^{i})\frac{d\theta}{\pi}~~;~~m=0,1 ~,
\end{equation}
\newline
which signifies that $B_{0}$ and $B_{1}$ can be obtained from
integrals involving the measured diffracted pressure field data
(for all angles $\theta$) on a circle of radius $b$. Once these
two coefficients are found, $\lambda^{1}$, $\mu^{1}$ and
$\rho^{1}$ can, in principle, be obtained from (\ref{8.7}), i.e.,
\newline
\begin{equation}\label{8.10}
B_{0}=\left[ \frac{\lambda+\mu-1}{\lambda+\mu}\right] ~~~,~~~
B_{1}=\left[ \frac{1-\rho}{1+\rho}\right] ~.
\end{equation}
wherein
\newline
\begin{equation}\label{8.10a}
\lambda:=\frac{\lambda^{1}}{\lambda^{0}}~~,~~ \mu
:=\frac{\mu^{1}}{\lambda^{0}}~~,~~\rho
:=\frac{\rho^{1}}{\rho^{0}}~.
\end{equation}
\newline
These relations:
\newline
\begin{itemize}
\item show that the field is
a {\it nonlinear} function of the material parameters, and
\item
apply equally-well to the  fluid model of the specimen (i.e.,
$\mu=0$).
\end{itemize}
$~$
\newline
Moreover, since only two pieces of data (i.e., $B_{0}$ and
$B_{1}$,  as expressed by (\ref{8.9})) are available in this
(second) order of approximation (in $\chi$), only one ($\rho$),
and a linear combination ($\lambda+\mu$) of the other two of the
three material parameters can be recovered via (\ref{8.10}).
\newline
\newline
More specifically, it is found that
\newline
\begin{equation}\label{8.11}
\lambda+\mu=\frac{1}{1-B_{0}}~~,~~\rho=\frac{1-B_{1}}{1+B_{1}}
 ~,
\end{equation}
\newline
which underlines the fact that the proposed technique enables:
\begin{itemize}
\item  an {\it explicit (partial) solution of the inverse problem} of the
reconstruction of $\rho$ and
$\frac{(c_{L}^{1})^{2}-(c_{T}^{1})^{2}}{(c_{L}^{0})^{2}}=\frac{\lambda+\mu}{\rho}$
when the material of the specimen is modeled by an elastic solid,
and
\item an {\it explicit (complete) solution of the inverse problem}
of the reconstruction of $\rho$ and
$\frac{(c_{L}^{1})^{2}}{(c_{L}^{0})^{2}}=\frac{\lambda}{\rho}$
when the material of the specimen is modeled by a fluid.
\end{itemize}
%
\section{A manner for obtaining the {\it complete} solution of
the inverse problem
for the elastic solid model of the specimen}
The previous analysis showed that $B_{n}=0~;~n=2,3,....$ to second
order in $\chi$, which is the reason why only two pieces of data
are available to reconstruct the three unknown material parameters
$\rho$, $\lambda$ and $\mu$.
\newline
\newline
It can be shown that by carrying out the perturbation analysis to
fourth order in $\chi$, not only $B_{0}$,  $B_{1}$, but also
$B_{2}$ are non-vanishing, so that by this means  one disposes of
the {\it three} pieces of data  necessary to reconstruct the three
unknown parameters.
\newline
\newline
The only difficulty with this procedure is that the relations
between $B_{0}$, $B_{1}$,  $B_{2}$ and $\rho$, $\lambda$, $\mu$
are much more complicated than previously, which fact makes it
impossible to obtain explicit algebraic expressions for $\rho$,
$\lambda$, $\mu$ in terms of $B_{0}$, $B_{1}$,  $B_{2}$.
\newline
\newline
It turns out that one is  faced with the problem of solving a
system of three nonlinear equations in three unknowns. The
procedure for solving this system is advantageously initialized
via the second-order-in-$\chi$ approximations of $\lambda+\mu$ and
$\rho$.
\section{Conclusion}
We have shown that the low-frequency perturbation scheme enables
an {\it explicit reconstruction} of two ($\lambda^{1}+\mu$ and
$\rho$) of the three material parameters $\rho$, $\lambda$, $\mu$
of a poroelastic cylindrical specimen, modeled as an equivalent
elastic solid circular cylinder.
\newline
\newline
Two single-frequency pieces of data, both involving the scattered
pressure field on a complete circle around the cylinder, are
required for this procedure.
\newline
\newline
Data obtained at other frequencies can be employed to reconstruct
the frequency-dependent complex parameter
$\lambda^{1}(\omega)+\mu(\omega)$ of cylindrical specimens modeled
as {\it viscoelastic} materials.
\newline
\newline
Above all, the same type of perturbation analysis can be employed
to obtain appropriate combinations of the Biot parameters when the
cylindrical specimen is modeled as a fully-poroelastic material
(in the sense of Biot).
\small
\bibliographystyle{unsrt}
\bibliography{biblio}

\begin{thebibliography}{10}

\bibitem{lapa84}
Langton C.M., Palmer S.B., and Porter R.W.
\newblock The measurement of broadband ultrasonic attenuation in cancellous
  bone.
\newblock {\em Engrg.in Med.}, 13:89--91, 1984.

\bibitem{layo86}
Lakes R., Yoon H.S., and Katz J.L.
\newblock Ultrasonic wave propagation and attenuation in wet bone.
\newblock {\em J.Biomed.Engrg.}, 8:143--148, 1986.

\bibitem{asco87}
Ashman~R. B., Corin~J. D., and Turner~C. H.
\newblock Elastic properties of cancellous bone: measurement by an ultrasonic
  technique.
\newblock {\em J.Biomech.}, 10:979--989, 1987.

\bibitem{wesa89}
Weaver R.L., Sachse W., and Niu L.
\newblock Transient ultrasonic waves in a viscoelastic plate: theory.
\newblock {\em J.Acoust.Am.}, 85:2255--2261, 1989.

\bibitem{mcpa91}
McKelvie T.J. and Palmer S.B.
\newblock The interaction of ultrasound with cancellous bone.
\newblock {\em Phys.Med.Biol.}, 36:1331--1340, 1991.

\bibitem{wi92}
Williams J.L.
\newblock Ultrasonic wave propagation in cancellous and cortical bone:
  prediction of some experimental results by \symbol{66}iot's theory.
\newblock {\em J.Acoust.Soc.Am.}, 91:1106--1112, 1992.

\bibitem{hoot97}
Hosokawa A. and Otani T.
\newblock Ultrasonic wave propagation in bovine cancellous bone.
\newblock {\em J. Acoust. Soc. Am.}, 101:558--562, 1997.

\bibitem{kiiy95b}
Kinra V.K. and Iyer V.R.
\newblock Ultrasonic measurement of the thickness, phase velocity, density or
  attenuation of a thin-viscoelastic plate. part ii: the inverse problem.
\newblock {\em Ultrasonics}, 33:110--121, 1995.

\bibitem{drbe98}
Droin P.~Berger G. and Laugier P.
\newblock Velocity dispersion of acoustic waves in cancellous bone.
\newblock {\em IEEE Trans.Ultrason.Ferroelect.Freq.Control}, 45:581--592, 1998.

\bibitem{we00}
Wear K.
\newblock Measurements of phase velocity and group velocity in human calcaneus.
\newblock {\em Ultrasound Med.Biol.}, 26:641--646, 2000.

\bibitem{fefe03a}
Fellah Z.E.A., Fellah M., Lauriks W., and Depollier C.
\newblock Direct and inverse scattering of transient acoustic waves by a slab
  of rigid porous material.
\newblock {\em J.Acoust.Soc.Am.}, 113:61--72, 2003.

\bibitem{bugi03}
Buchanan J.L., Gilbert R.P., Wirgin A., and Xu~Y.
\newblock Transient reflection and transmission of ultrasonic waves in
  cancellous bones.
\newblock {\em Appl.Math.Computation}, 142:561--573, 2003.

\bibitem{femi03}
Fellah Z.E.A., Mitri F.G.and~Depollier C., Berger S., Lauriks W., and Chapelon
  J.-Y.
\newblock Characterization of porous materials with a rigid frame via reflected
  waves.
\newblock {\em J.Appl.Phys.}, 94:7914--7922, 2003.

\bibitem{febe04d}
Fellah Z.E.A., Berger, Lauriks W., Depollier C., Aristegui C., and Chapelon
  J.Y.
\newblock Measuring the porosity and the tortuosity of porous materials via
  reflected waves at oblique incidence.
\newblock {\em J.Acoust.Soc.Am.}, 113:2424--2433, 2004.

\bibitem{bugi04}
Buchanan J.L., Gilbert R.P., and Khashanah K.
\newblock Determination of the parameters of cancellous bone using low
  frequency acoustic measurements.
\newblock {\em J.Comput.Acoust.}, 12:99--126, 2004.

\bibitem{khvi81}
Khalil T.B., Viano D.C., and Taber L.A.
\newblock Vibrational characteristics of the embalmed human femur.
\newblock {\em J. Sound Vibr.}, 75:417--436, 1981.

\bibitem{zale04}
Zadler B.J., Le~Rousseau J.H.L., Scales J.A., and Smith M.L.
\newblock Resonant ultrasound spectroscopy: theory and application.
\newblock {\em Geophys.J.Int.}, 156:154--169, 2004.

\bibitem{gusc99}
Gurevich B. and Schoenberg M.
\newblock Interface conditions for \symbol{66}iot's equations of
  poroelasticity.
\newblock {\em J.Acoust.Soc.Am.}, 105:2585--2589, 1999.

\bibitem{ya83}
Yamamoto T.
\newblock Acoustic propagation in the ocean with a poro-elastic bottom.
\newblock {\em J.Acoust.Soc.Am.}, 73:1587--1596, 1983.

\bibitem{ch95}
Chotiros N.P.
\newblock Biot model of sound propagation in water-saturated sand.
\newblock {\em J.Acoust.Soc.Am.}, 97:199--214, 1995.

\bibitem{be81}
Berryman J.G.
\newblock Elastic wave propagation in fluid-saturated porous media.
\newblock {\em J.Acoust.Soc.Am.}, 69:416--424, 1981.

\bibitem{stka81}
Stoll R.D. and Kan T.-K.
\newblock Reflection of acoustic waves at a water-sediment interface.
\newblock {\em J.Acoust.Soc.Am.}, 70:149--156, 1981.

\bibitem{bugi98}
Buchanan J.L. and Gilbert R.P.
\newblock Transmission loss in a shallow ocean over a two-layer seabed.
\newblock {\em Int.J.Solids Struc.}, 35:4779--4801, 1998.

\bibitem{huel90}
Hughes S., Ellis D., Chapman D., and Stahl P.
\newblock Low-frequency acoustic propagation loss in shallow water over
  hard-rock seabeds covered by a thin layer of elastic-solid sediment.
\newblock {\em J. Acoust. Soc. Am.}, 88:283--297, 1990.

\bibitem{moba96}
Morochnik V. and Bardet J.P.
\newblock Viscoelastic approximation of poroelastic media for wave scattering
  problems.
\newblock {\em Soil Dynam.Earthqu.Engrg.}, 15:337--346, 1996.

\bibitem{vi80}
Vidmar P.
\newblock The effect of sediment rigidity on bottom reflection loss in a
  typical deep sea sediment.
\newblock {\em J. Acoust. Soc. Am.}, 68:634--648, Sep 1980.

\bibitem{deal88}
Depollier C., Allard J.F., and Lauriks W.
\newblock Biot theory and stress-strain equations in porous sound-absorbing
  materials.
\newblock {\em J.Acoust.Soc.Am.}, 84:2277--2279, 1988.

\end{thebibliography}
\end{document}